\documentclass[aps,prl,twocolumn,showpacs,floatfix,superscriptaddress]{revtex4-1}

\usepackage{times,mathrsfs}
\usepackage{graphicx,color}
\usepackage{amsmath,amssymb}
\usepackage[colorlinks,citecolor=red]{hyperref}
\usepackage[normalem]{ulem}

\begin{document}

\newcommand\cut{\text{cut}}
\newcommand\MF{\text{MF}}
\newcommand\eq{\text{eq}}
\newcommand\fwd{\text{fwd}}
\newcommand\bwd{\text{bwd}}
\newcommand\erfc{\operatorname{erfc}}

\newcommand\hatH{{\hat{H}}}
\newcommand\hatS{{\hat{S}}}
\newcommand\hatW{{\hat{W}}}

\newcommand\calH{{\mathcal{H}}}
\newcommand\calI{{\mathcal{I}}}
\newcommand\calA{{\mathcal{A}}}
\newcommand\calB{{\mathcal{B}}}
\newcommand\calC{{\mathcal{C}}}
\newcommand\calX{{\mathcal{X}}}
\newcommand\calY{{\mathcal{Y}}}

\newcommand\varA{{\mathscr{A}}}
\newcommand\varO{{\mathscr{O}}}

\newcommand\bfS{\mathbf{S}}
\newcommand\BfZ{\mathbb{Z}}

\newcommand\half{\frac{1}{2}}

\newcommand\avg[1]{\mathinner{\langle{\textstyle#1}\rangle}}
\newcommand\set[1]{\mathinner{\{\textstyle#1}\}}

\newcommand\marginnote[1]{\marginpar{\tiny\color{red}#1}}
\newcommand\red{\color{red}}
\newcommand\gray{\color{gray}}
\newcommand\blue{\color{blue}}
\newcommand{\revgw}[1]{{\color{blue}#1}}
\newcommand{\revjj}[1]{{\color{magenta}#1}}
\newcommand{\revpv}[1]{{\color{green}#1}}
\definecolor{gray}{gray}{0.5}

\title{Emergent Universality in Nonequilibrium Processes of Critical Systems}

\author{Danh-Tai Hoang}
\affiliation{Asia Pacific Center for Theoretical Physics (APCTP),
Pohang, Gyeongbuk 790-784, Korea}
\affiliation{Department of Science, Quang Binh University, Dong Hoi, Quang Binh 510000, Vietnam}

\author{B. Prasanna Venkatesh}
\affiliation{Asia Pacific Center for Theoretical Physics (APCTP),
Pohang, Gyeongbuk 790-784, Korea}
\affiliation{Institute for Quantum Optics and Quantum Information of the Austrian Academy of Sciences, Technikerstra\ss e 21a, Innsbruck 6020, Austria}

\author{Seungju Han}
\affiliation{%
  Department of Physics, Korea University, Seoul 136-701, Korea}

\author{Junghyo Jo}
\affiliation{Asia Pacific Center for Theoretical Physics (APCTP),
Pohang, Gyeongbuk 790-784, Korea}
\affiliation{Department of Physics, Pohang University of Science and Technology (POSTECH), Pohang, Gyeongbuk 790-784, Korea}

\author{Gentaro Watanabe}
\affiliation{Asia Pacific Center for Theoretical Physics (APCTP),
Pohang, Gyeongbuk 790-784, Korea}
\affiliation{Center for Theoretical Physics of Complex Systems, Institute for Basic Science (IBS), Daejeon 305-732, Korea}

\author{Mahn-Soo Choi}
\email{choims@korea.ac.kr}
\affiliation{%
  Department of Physics, Korea University, Seoul 136-701, Korea}

\date{\today}

\begin{abstract}
We examine the Jarzynski equality for a quenching process across the critical point of second-order phase transitions, where absolute irreversibility and the effect of finite-sampling of the initial equilibrium distribution arise on an equal footing. We consider the Ising model as a prototypical example for spontaneous symmetry breaking and take into account the finite sampling issue by introducing a tolerance parameter. For a given tolerance parameter, the deviation from the Jarzynski equality depends onthe reduced coupling constant and the system size. In this work, we show that the deviation from the Jarzynski equality exhibits a universal scaling behavior inherited from the critical scaling laws of second-order phase transitions.
\end{abstract}


\maketitle

\paragraph{Introduction ---}
Fluctuation theorems (FTs) provide universal and exact relations for nonequilibrium processes irrespective of how far a system is driven away from equilibrium.
The discovery of FTs is a major development in nonequilibrium statistical mechanics, pioneered by Bochkov and Kuzovlev \cite{Bochkov77a,*Bochkov77b, Bochkov79a,*Bochkov79b} for a special case and thriving with the celebrated equalities of Jarzynski \cite{Jarzynski97a} and Crooks \cite{Crooks99a} which hold for general forcing protocols (see, e.g., \cite{Jarzynski08rev, Jarzynski11rev, Pitaevskii11rev, Seifert12rev} and references therein for recent reviews).

Since the discoveries of the Jarzynski equality (JE) and the Crooks relation, a large effort has been made to find applications of these universal relations. As a representative example, FTs provide a unique way to evaluate the free energy difference $\Delta F$ between equilibrium states through nonequilibrium processes \cite{Jarzynski97a}, which could be useful for systems such as complex molecules \cite{Hummer01a, Liphardt02a} that take a very long time to reach an equilibrium state.
FTs have also been exploited to study the non-equilibrium dynamics \cite{Silva08b,Paraan09a,Palmai14a,Dutta15a}, to show the emergence of thermodynamics out of microscopic reversibility \cite{Dorner12a}, and to investigate the universal behaviors of the work-distribution tails \cite{Gambassi12a} in quantum critical systems.
Further, FTs by themselves serve as useful formulae which simplify theoretical derivations and facilitate important developments such as information thermodynamics \cite{Sagawa08a}.

Although the FTs hold universally, they require sufficient sampling from the states of the initial ensemble, which leads to a \emph{convergence problem} in many situations \cite{Zuckerman02a, Gore03a, Jarzynski06a, Palassini11a, Suarez12a}.
For example, consider the JE, $\avg{e^{-\sigma}}=1$, where
$\sigma=\beta(W-\Delta{F})$ is the irreversible entropy production, $W$ the
work performed to the system, and $\beta$ the inverse temperature. The
realizations of a thermodynamic process which yield the dominant contribution
to the ensemble average of $e^{-\sigma}$ can be very different from typical
realizations under the same condition.
In such a situation, sufficient sampling of the dominant realizations becomes intractable with increasing system size, and in reality the JE is hard to verify to high accuracy with a finite number of samples in the ensemble.

\begin{figure}
\centering
\includegraphics[width=7cm]{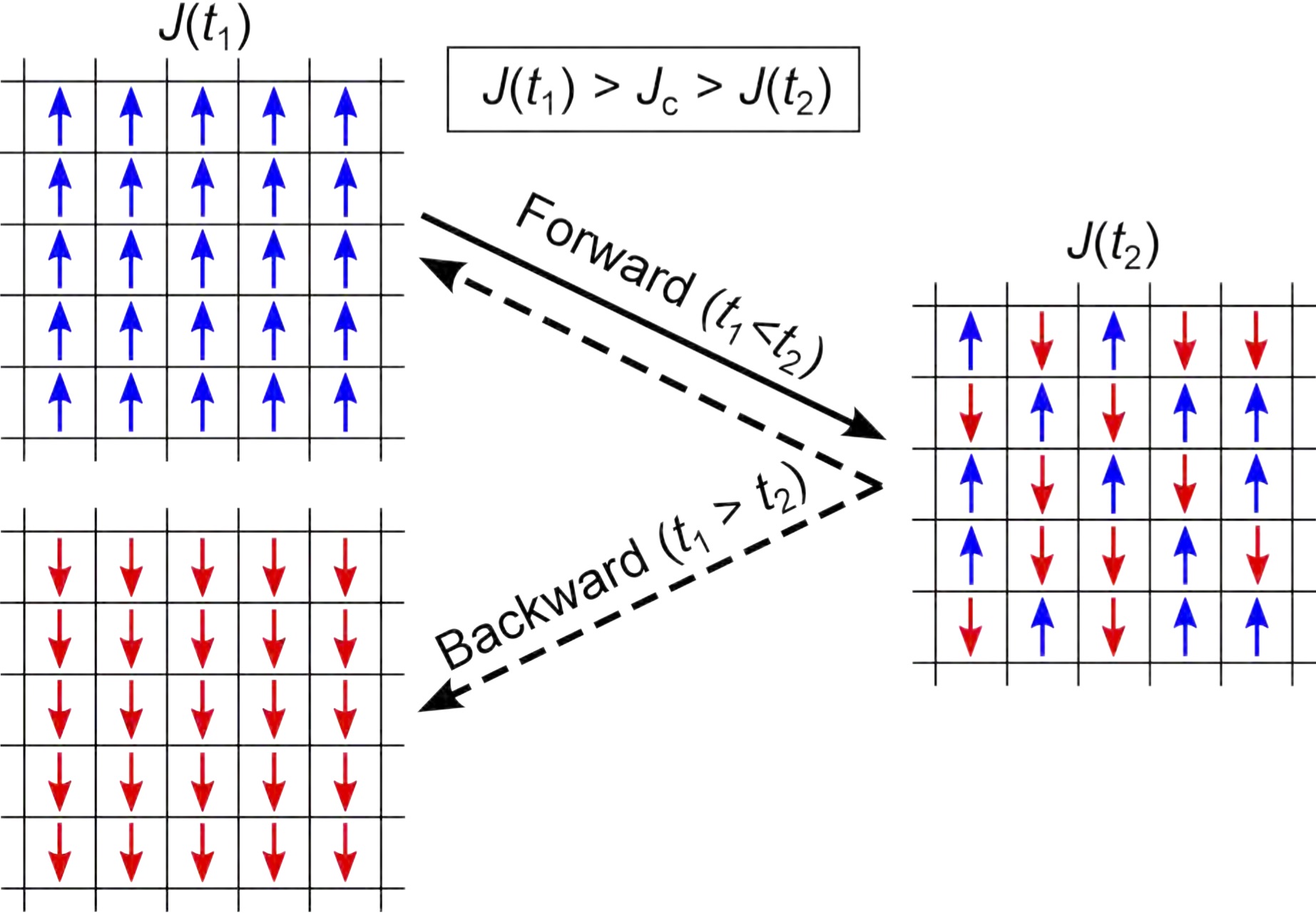}
\caption{Schematic representation of the absolute irreversibility in the quench dynamics of Ising model. In the forward process, the system is initially at equilibrium with positive spontaneous magnetization, whereas in the backward process the initial equilibrium state is without spontaneous magnetization. When the coupling strength between the spins is increased across the critical point (from the right to left in the figure), the system can have either positive or negative spontaneous magnetization. The latter case has no corresponding forward path, which results in the absolute irreversibility.}
\label{Paper::fig:1}
\end{figure}

Moreover, even in the ideal case with sufficient sampling, there are a class of processes such as the free expansion of a gas, to which the JE does not apply due to a fundamental reason that has been referred to as {\it absolute irreversibility} \cite{Lua05a, Sung05a, Gross05a, Jarzynski05a, Horowitz10a, Murashita14a}. A process is called absolutely irreversible if there exists a path in phase space whose probability to occur in the forward direction is zero while that in the reverse direction is nonzero, or vice versa. A typical situation occurs when the accessible phase spaces for the system at the beginning and end of a protocol are not identical. This is indeed the case for the free expansion of initially confined particles whose accessible phase space is increased by removing the partitioning barrier.


In this Letter, we explore the fact that
in systems driven through
second-order phase transitions, both the absolute irreversibility and the
convergence issue can take place on an equal footing.
Using numerical simulations and the scaling theory of phase transitions, the deviation from the JE is examined as a function of the system size and the reduced coupling constant.
It exhibits a universal scaling behavior inherited from the critical scaling of the correlation length and the relaxation time in second-order phase transitions.
This finding may provide a unique application of the FTs to study the dynamical properties of phase transitions.

While the detailed arguments and analyses are discussed below, the essential points can be summarized as follows: On the one hand, a natural partitioning of the phase space emerges as a consequence of the ergodicity breaking in the ordered phase \cite{Goldenfeld92a} in contrast to the partitioning externally imposed in the example of free expansion.
The resultant absolute irreversibility is illustrated in Fig.~\ref{Paper::fig:1} for the  Ising model, which as the simplest model showing spontaneous symmetry breaking will be used to anchor the rest of our discussions. It is expected (and proven rigorously in Section II of \cite{Supplement}) that the spontaneous breaking of $\BfZ_2$ symmetry leads to a doubling of the accessible phase space resulting in $\avg{e^{-\sigma}}=1/2$
when the system is quenched from the equilibrium ordered phase to the disordered side.
On the other hand, in such a process the configurations with vanishing (spatial) mean order parameter give major contributions to $\avg{e^{-\sigma}}$, while such configurations are extremely rare in the initial equilibrium in the ordered phase.
An observation over a finite time in realistic experiments and numerical simulations inevitably leads to insufficient sampling. In this work, we account for such insufficient sampling by introducing a tolerance parameter to neglect some unlikely configurations in the initial equilibrium distributions. It is intriguing that the deviation from the JE is determined universally for a given finite value of this tolerance.

\paragraph{Model and Notations ---}

We take the Ising model as the simplest example showing spontaneous symmetry breaking. It consists of $N=L^d$ spins $\{S_j=\pm 1|j=1,\cdots,N\}$ on a $d$-dimensional lattice of lateral size $L$ whose interaction is governed by the Hamiltonian
\begin{equation}
\label{Paper::eq:8}
\beta{H} = -J\sum_{\langle ij \rangle}S_iS_j \,,
\end{equation}
where $\langle ij\rangle$ denote a pairs of nearest-neighbor sites and $J$ the coupling strength. We represent a spin configuration (microstate) by
\begin{math}
\bfS = (S_1,S_2,\cdots,S_N),
\end{math}
its magnetization per spin by $S\equiv N^{-1}\sum_jS_j$,
and the set of all spin configurations by $\calH$.
The configuration space $\calH$ consists of $\calH_\pm$ and $\calH_0$,
\begin{math}
\calH \equiv \calH_+\cup\calH_-\cup\calH_0,
\end{math}
where
\begin{equation}
\calH_\pm \equiv \set{\bfS \;|\; S \gtrless 0} \,,\quad
\calH_0 \equiv \set{\bfS \;|\; S = 0} \,.
\end{equation}
$\calH_0$ is naturally empty for odd $N$. For even $N$, the contribution of $\calH_0$ is negligible (probability measure is zero) in the thermodynamic limit and hereafter we ignore it.

We consider, for simplicity and to confirm with usual protocols discussed in the context of JE, quenching processes where the coupling constant $J(t)$ varies while temperature is kept constant. As usual, we define the \emph{reduced coupling constant} by
\begin{math}
\epsilon(t) = 1-J(t)/J_c,
\end{math}
where $J_c$ is the value at the critical point.
As the time $t$ changes from $t_i$ to $t_f$, the reduced coupling $\epsilon$ changes from $\epsilon_i\equiv\epsilon(t_i)$ to $\epsilon_f\equiv\epsilon(t_f)$ and the Hamiltonian from $H_i\equiv H(t_i)$ to $H_f\equiv H(t_f)$. In order to discuss absolute irreversibility, we will be mostly interested in  quenching from the ordered ($\epsilon_i<0$) to disordered ($\epsilon_f>0$) phase. To avoid unnecessary complications, unless specified otherwise, we will consider symmetric quenching: $\epsilon_f=-\epsilon_i=\epsilon_0>0$.

Although the quenching process drives the system out of equilibrium, many physical effects and quantities are still described in terms of the initial and final \emph{equilibrium} distribution functions
\begin{equation}
\rho_{i/f}(\bfS) = Z_{i/f}^{-1}\, e^{-\beta{H}_{i/f}(\bfS)} \,,
\end{equation}
where $Z_{i/f}$ are the respective partition functions. Since we start from the ordered phase at initial time, the allowed spin configurations are restricted either to $\calH_+$ or $\calH_-$ due to spontaneous symmetry breaking. For keeping the discussion specific we take the spin configurations to be in $\calH_+$ giving the initial partition function
\begin{math}
Z_i = \sum_{\bfS\in\calH_+}e^{-\beta{H}_i}
\end{math}
while the final partition function is given by
\begin{math}
Z_f = \sum_{\bfS\in\calH}e^{-\beta{H}_f}
\end{math}
as usual. We will see that the restriction of the initial spin configurations has vital consequences.

It turns out that the equilibrium probabilities $P_{i/f}(S)$ of magnetization per spin $S$ are particularly useful, and related to $\rho_{i/f}(\bfS)$ by
\begin{equation}
P_{i/f}(S)
= \sum_{\bfS}\delta(NS-{\textstyle\sum_j}S_j)
\rho_{i/f}(\bfS).
\end{equation}

\begin{figure}
\centering
\includegraphics[width=6cm]{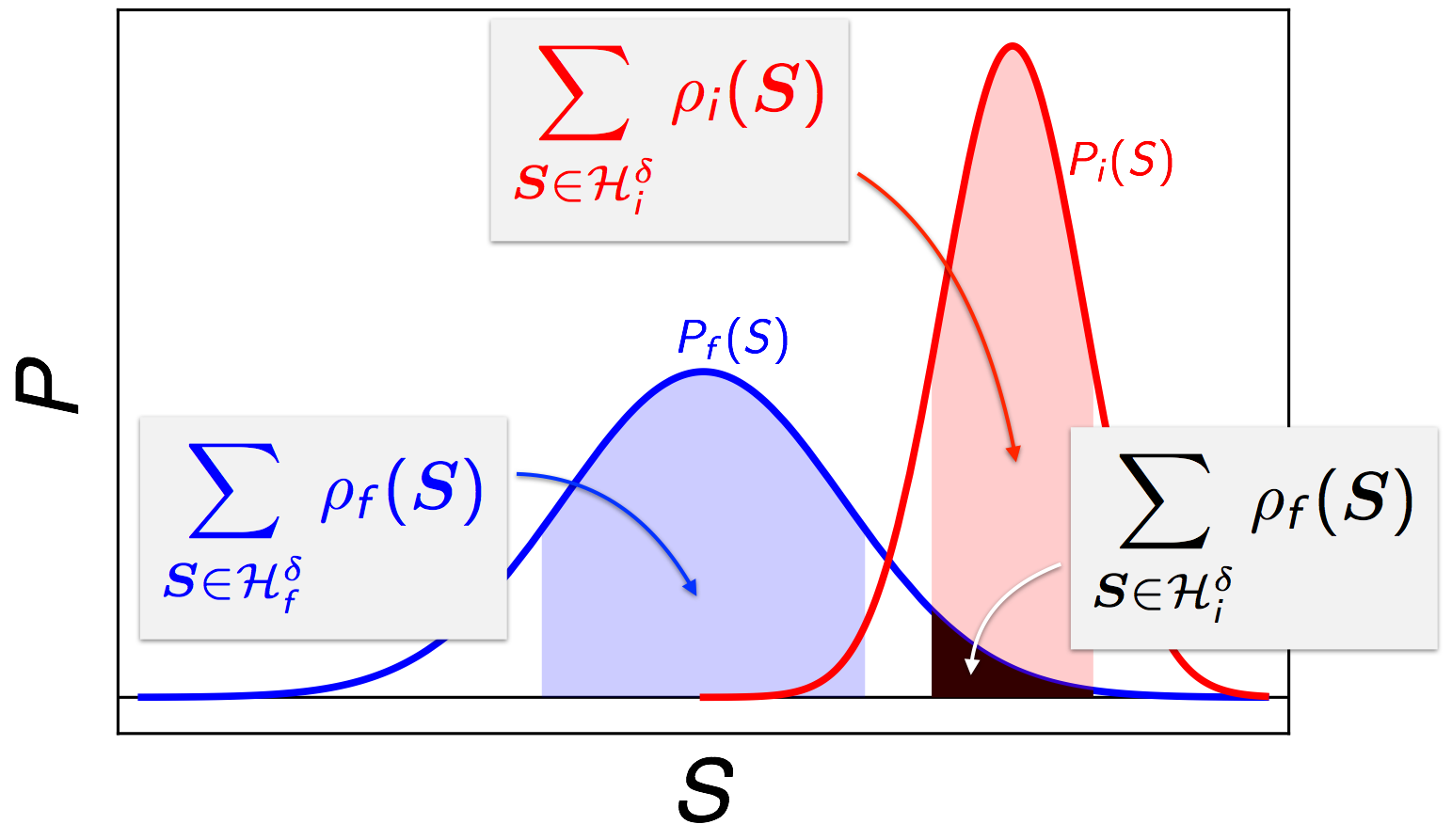}
\caption{Schematic representation of the sets $H_{f/i}^\delta$ of allowed spin configurations and their relations to $\avg{e^{-\sigma}}_\delta$. For a given tolerance $\delta$, $\avg{e^{-\sigma}}_\delta$ is given by the ratio of the areas in black and blue shade.
}
\label{Paper::fig:2}
\end{figure}

\paragraph{Tolerance parameter ---}
In the so-called ``sudden'' (infinitely fast) quenching (more general cases are discussed in \cite{Supplement}), the system does not have enough time to change its distribution over spin configurations, and hence the initial equilibrium distribution is preserved throughout the whole process. The work distribution is thus completely determined by $\rho_i$, leading to
\begin{equation}
\avg{e^{-\beta W}}
= \avg{e^{-\beta(H_f-H_i)}}_{\rho_i} \,,
\end{equation}
where the average $\avg{\cdots}_{\rho_i}$ is over the initial distribution $\rho_i(\bfS)$.
Recall that the initial spin configurations are restricted to $\calH_+$ due to spontaneous symmetry breaking. The exponential average of the entropy production $\sigma=\beta(W-\Delta{F})$ follows easily from $\langle e^{-\beta W} \rangle$ by multiplying by the exponential of the free energy change given by $\beta\Delta{F}=-\log(Z_f/Z_i)$.
In realistic experiments and numerical simulations, 
spin configurations with exponentially small probability do not take actual effects. Therefore it is natural to ignore such spin configurations up to certain tolerance $\delta$. Specifically, for a given probability distribution $\rho$ and tolerance $\delta$, we define the set of kept spin configurations $\calH(\rho,\delta)$ and the cutoff probability $\rho_\cut(\rho,\delta)$ by the following two conditions (see also Fig.~\ref{Paper::fig:2}):
\begin{subequations}
\label{Paper::eq:1}
\begin{gather}
\calH(\rho,\delta) = \set{\bfS\;|\;\rho(\bfS)>\rho_\cut(\rho,\delta)} \,,\\
\sum_{\bfS\in\calH(\rho,\delta)}\rho(\bfS) = 1 - \delta.
\end{gather}
\end{subequations}
We introduce the short-hand notations $\calH_{i/f}^\delta\equiv\calH(\rho_{i/f},\delta)$. The corresponding partition functions are given by
\begin{math}
Z_{i/f}^\delta = (1-\delta)Z_{i/f}.
\end{math}
With a finite tolerance $\delta$, the average $\avg{e^{-\beta W}}_\delta$ is given by 
\begin{equation}
\avg{e^{-\beta W}}_\delta
= \sum_{\bfS\in\calH_i^\delta} e^{-\beta(H_f-H_i)}
\frac{e^{-\beta{H}_i}}{Z_i^\delta}
= \sum_{\bfS\in\calH_i^\delta}
\frac{e^{-\beta{H}_f}}{Z_i^\delta} \,.
\end{equation}
The free energy change is not affected by tolerance, $\beta\Delta{F}=-\log(Z_f^\delta/Z_i^\delta)=-\log(Z_f/Z_i)$. We thus obtain
\begin{equation}
\label{Paper::eq:10}
\avg{e^{-\sigma}}_\delta
= \frac{\sum_{\bfS\in\calH_i^\delta}\rho_f(\bfS)}
{\sum_{\bfS\in\calH_f^\delta}\rho_f(\bfS)}.
\end{equation}
Equation~(\ref{Paper::eq:10}) is one of our main results and manifests several features to be stressed: (i) As illustrated schematically in Fig.~\ref{Paper::fig:2}, $\avg{e^{-\sigma}}_\delta$ depends crucially on the overlap of $\calH_i^\delta$ and $\calH_f^\delta$. For finite $\delta$, well separated initial and final distributions lead to vanishing $\avg{e^{-\sigma}}_\delta$. For $\delta=0$, on the other hand, $\calH_i^\delta=\calH_+$ and $\calH_f^\delta=\calH$, and hence $\langle e^{-\sigma}\rangle = 1/2$ validating the heuristic analysis presented in Fig.~\ref{Paper::fig:1}. (ii) Equation~(\ref{Paper::eq:10}) also demonstrates how the convergence issue arises in quenching process of phase transitions. Namely, the dominant contributions to the ensemble average of $e^{-\sigma}$ comes from the spin configurations with larger $\rho_f(\bfS)$ whereas the initial equilibrium is governed by those with larger $\rho_i(\bfS)$.
(iii) Equation~(\ref{Paper::eq:10}) describes highly \emph{non-equilibrium} processes merely in terms of \emph{equilibrium} distributions, a remarkably simple way to study $\avg{e^{-\sigma}}_\delta$.

The tolerance scheme~(\ref{Paper::eq:1}) in terms of the \emph{microscopic} spin configurations $\bfS$ is still difficult to implement in practice. For example, the tolerance parameter $\delta$ corresponding to the actual finite sampling is unknown or very difficult to estimate in most cases. For this reason, we introduce another \emph{operational} tolerance scheme in terms of the \emph{macroscopic} order parameter $S$: Given $P(S)$, we define the interval of relevant magnetization $\calI(P,\delta)$ and the cutoff $P_\text{cut}(P,\delta)$ by
\begin{subequations}
\label{Paper::eq:2}
\begin{gather}
\calI(P,\delta) = \set{S\;|\;P(S)>P_\text{cut}(P,\delta)} \,, \\
\int_{\calI(P,\delta)}{dS}\,P(S) = 1-\delta.
\end{gather}
\end{subequations}
Note that the relation~(\ref{Paper::eq:10}) does not depend on a particular tolerance scheme, and for the scheme~(\ref{Paper::eq:2}) it reads as
\begin{equation}
\label{Paper::eq:4}
\avg{e^{-\sigma}}_\delta
= \frac{\int_{\calI_i^\delta}{dS}\,P_f(S)}{
  \int_{\calI_f^\delta}{dS}\,P_f(S)},
\end{equation}
where $\calI_{i/f}^\delta\equiv\calI(P_{i/f},\delta)$. Below we will mainly use the tolerance scheme~(\ref{Paper::eq:2}).

\paragraph{Scaling Behavior ---}
We now examine $\avg{e^{-\sigma}}_\delta$ in Eq.~(\ref{Paper::eq:4}) more closely, focusing on its scaling behavior inherited from the spontaneous symmetry breaking.

\begin{figure}
\centering
\includegraphics[width=4cm]{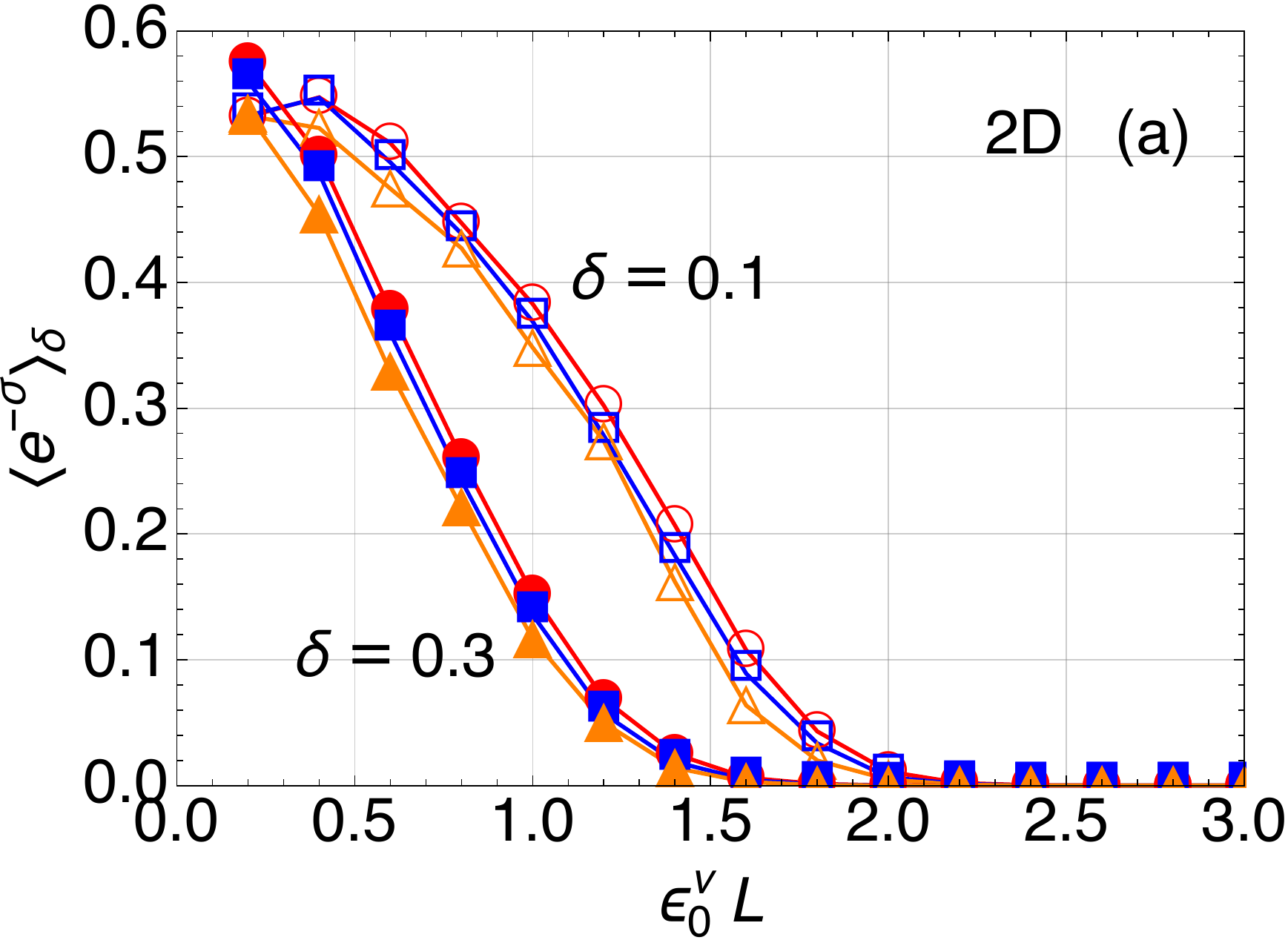}
\includegraphics[width=4cm]{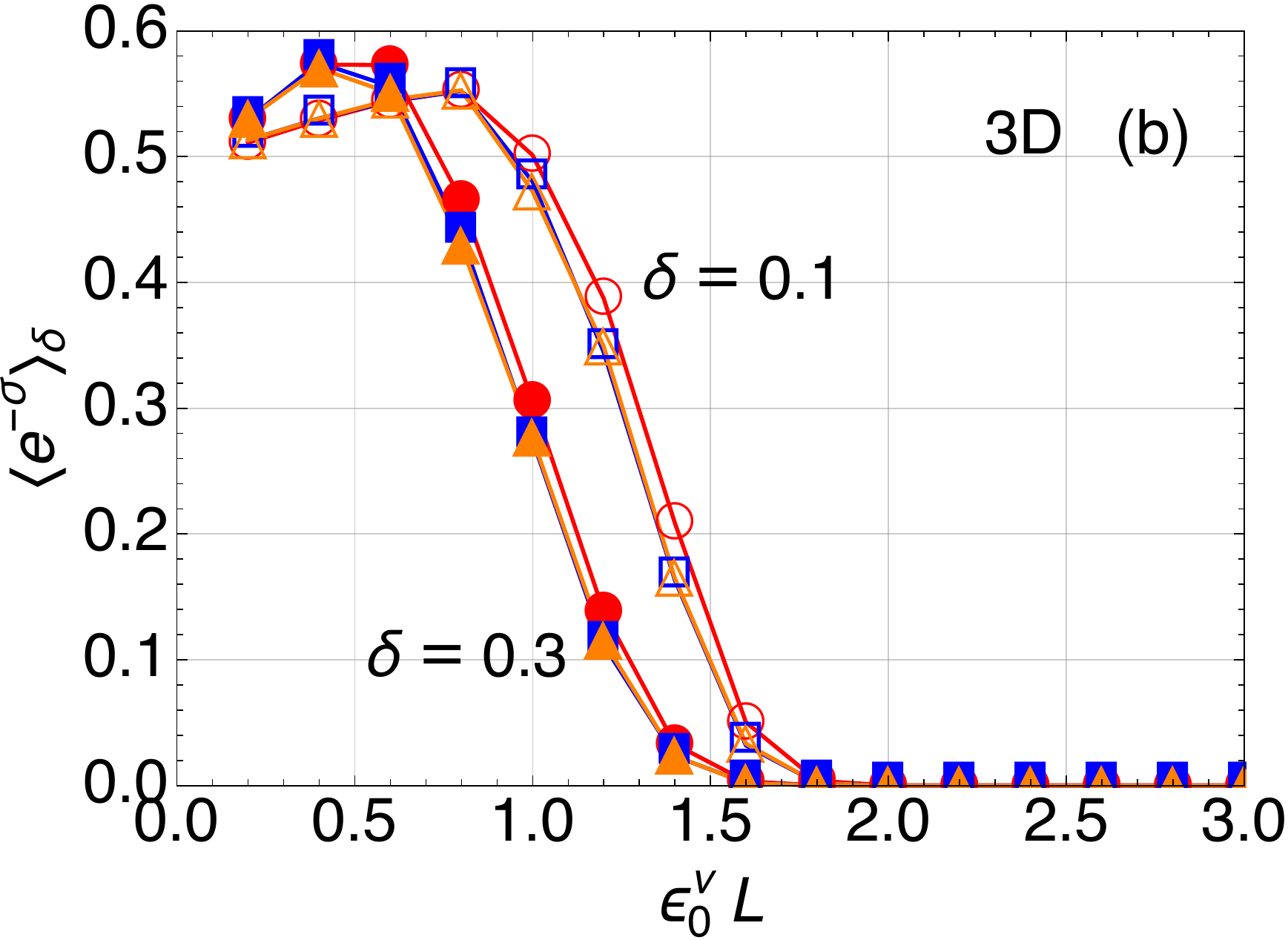}
\includegraphics[height=37mm]{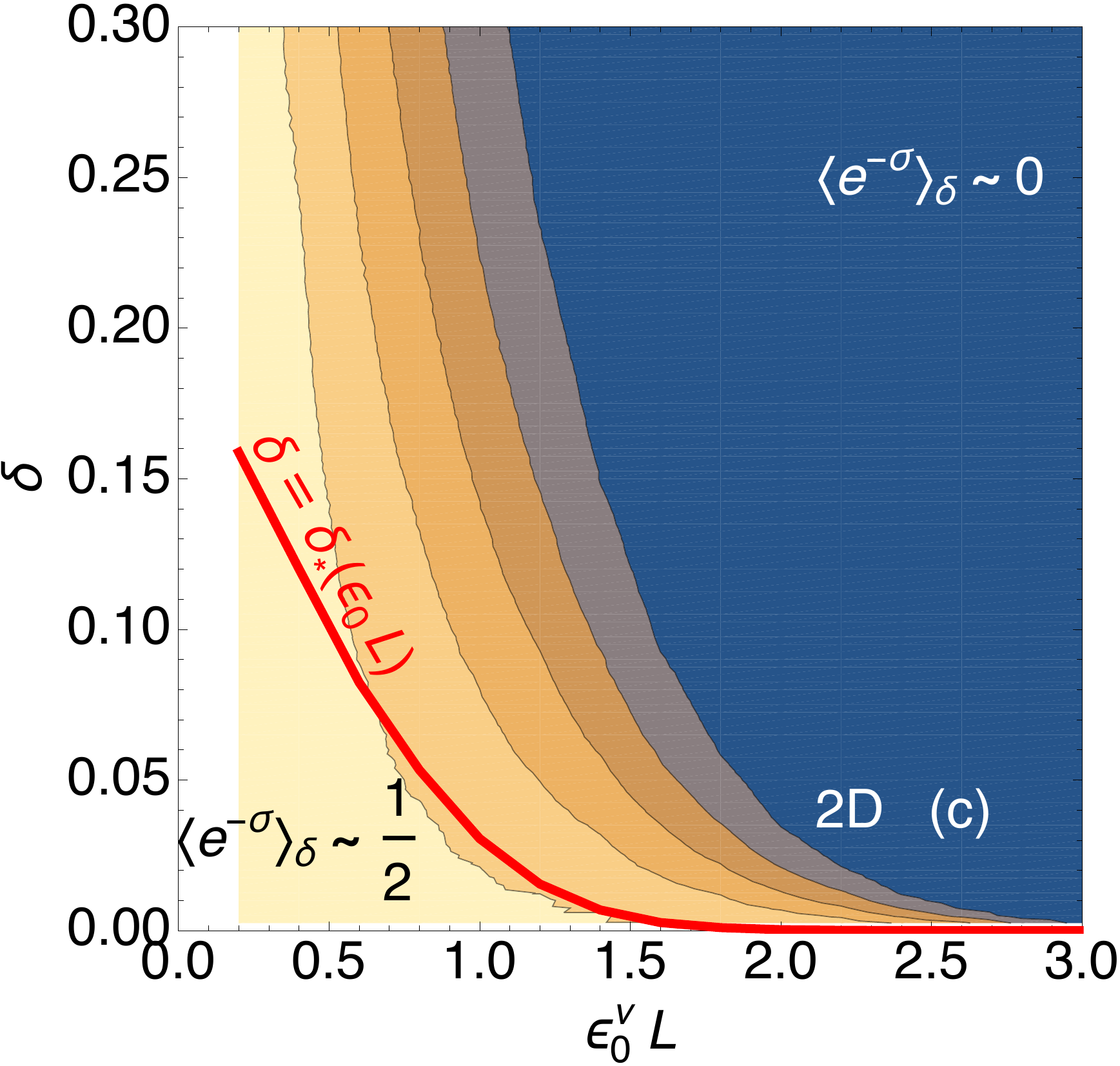}
\includegraphics[height=37mm]{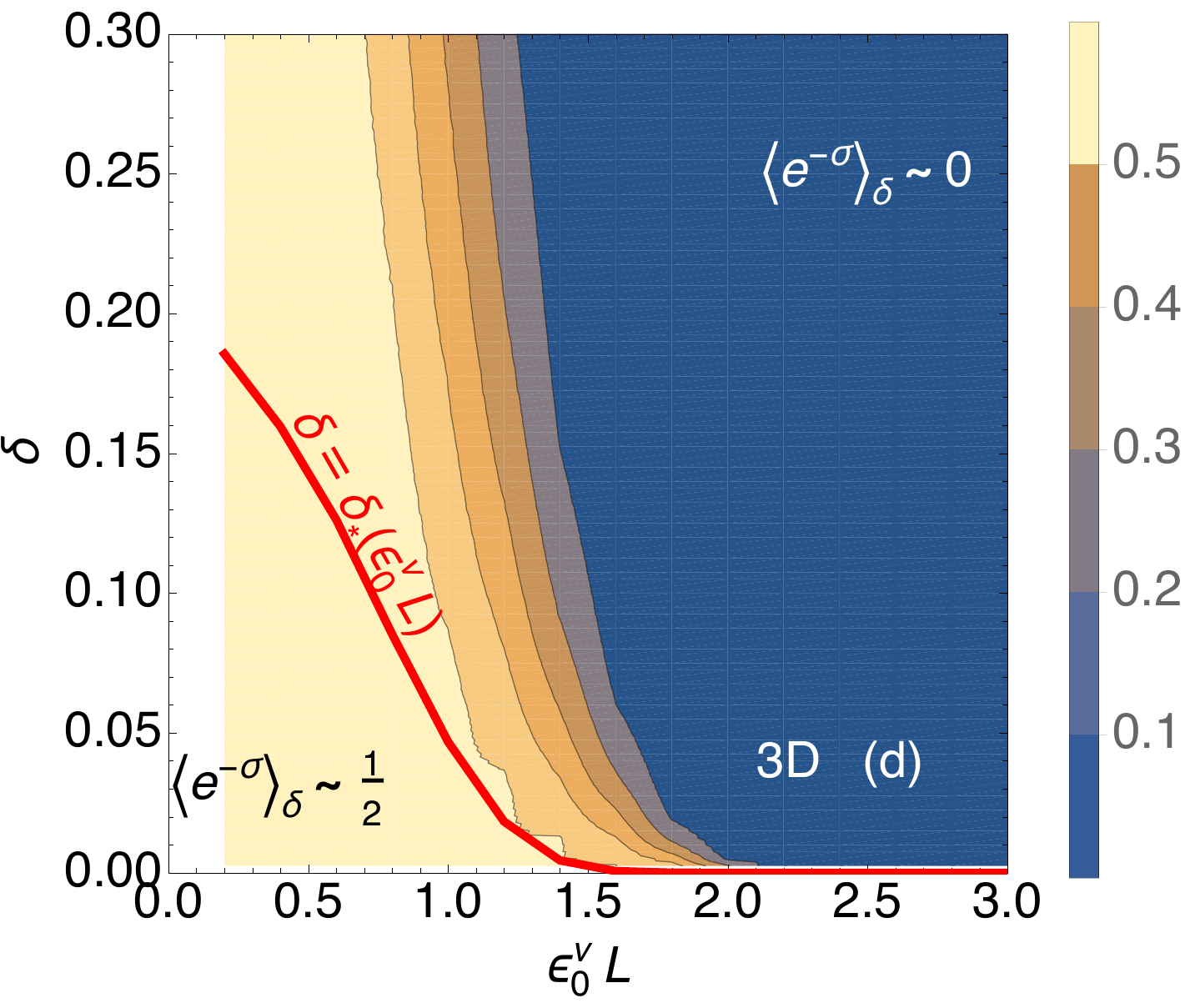}
\caption{$\avg{e^{-\sigma}}_\delta$ from Monte Carlo simulations of the Ising model on a 2D square (a,c) and 3D cubic (b,d) lattice. (a,b) Plots of $\avg{e^{-\sigma}}_\delta$ as a function of $\epsilon_0^\nu L$ for $\delta=0.1$ (empty symbols) and $\delta=0.3$ (filled symbols). $L=50$ (circles), $L=100$ (squares) and $L=200$ (triangles) in 2D (a) and $L=20$ (circles), $L=40$ (squares) and $L=50$ (triangles) in 3D (b). Note that, for a given $\delta$, all the curves for different $L$ collapse into a single curve.
  (c),(d) The contour plot of $\avg{e^{-\sigma}}_\delta$ as a function of $\epsilon_0^\nu L$ and $\delta$ \cite[Section~I.B]{Supplement}. The thick red line represents the crossover boundary, $\delta=\delta_*(\epsilon_0^\nu L)$.}
\label{Paper::fig:3}
\end{figure}

Before providing detailed scaling analyses below, we first summarize in Fig.~\ref{Paper::fig:3} the behavior of $\avg{e^{-\sigma}}_\delta$  as a  function of $\epsilon_0^\nu L$ and $\delta$, where $\nu$ is the critical exponent of the correlation length, $\xi \sim |\epsilon|^{-\nu}$. In Fig.~\ref{Paper::fig:3} we have performed Monte Carlo simulations \cite{Supplement} of the Ising model on two-dimensional (2D) square and three-dimensional (3D) cubic lattices. We have calculated the distributions $P_{i/f}(S)$ and then $\avg{e^{-\sigma}}_\delta$ based on Eq.~(\ref{Paper::eq:4}). Figure~\ref{Paper::fig:3} demonstrates three remarkable features:
First, for a given $\delta$, $\avg{e^{-\sigma}}_\delta$ is a universal function of $\epsilon_0^\nu L$ only ($\nu=1$ in 2D and $\nu=0.6301$ in 3D~\cite{Matz94a}) and does not depend separately on $\epsilon_0$ and $L$; see Fig.~\ref{Paper::fig:3} (a) and (b). The discovery of this universality is another one of our main results.
Second, in the parameter space of $\epsilon_0^\nu L$ and $\delta$, $\avg{e^{-\sigma}}_\delta\sim 0$ in the limit of $(\epsilon_0^\nu L,\delta)\to(\infty,1)$ while $\avg{e^{-\sigma}}_\delta\sim 1/2$ in the opposite limit $(\epsilon_0^\nu L,\delta)\to(0,0)$; see Fig.~\ref{Paper::fig:3} (c) and (d).
Third, for fixed $\delta$, it is suppressed exponentially, $\avg{e^{-\sigma}}_\delta\sim e^{-(\epsilon_0^\nu L)^d/2}$, for sufficiently large systems ($\epsilon_0^\nu L\gg 1$) while  it recovers $\avg{e^{-\sigma}}_\delta\simeq1/2$~\cite{endnote:5} for small systems ($\epsilon_0^\nu L\lesssim 1$).

According to Eq.~(\ref{Paper::eq:4}), the overlap between the distribution functions $P_{i/f}(S)$ plays a crucial role in $\avg{e^{-\sigma}}_\delta$. Let us investigate this overlap based on scaling analysis. For sufficiently large systems, the distributions are rather sharp and it suffices to characterize them by the peaks and their widths.
The initial distribution $P_i(S)$ at $\epsilon=-\epsilon_0$ has a peak near $S=M_i\equiv\avg{S}_{\rho_i}$ of width $\Delta_i=\sqrt{\avg{S^2}_{\rho_i}-\avg{S}_{\rho_i}^2}$. According to the fluctuation-dissipation theorem~\cite{Goldenfeld92a}, $\Delta_i$ is related to the equilibrium susceptibility $\chi_i$ by
$\Delta_i = \sqrt{\chi_i/L^d}$.
Similarly, the final distribution $P_f(S)$ at $\epsilon=+\epsilon_0$ has a peak near $S=M_f\equiv\avg{S}_{\rho_f}=0$ of width $\Delta_f= \sqrt{\chi_f/L^d}$.
The magnetization and susceptibility satisfy the standard scaling behaviors:
\begin{align}
\label{Paper::eq:3}
M_i(\epsilon_0,L)
&\sim \epsilon_0^\beta\Phi_M(\epsilon_0^\nu L) \,, \\
\label{Paper::eq:5}
\chi_i(-\epsilon_0,L)\sim \chi_f(\epsilon_0,L)
&\sim \epsilon_0^{-\gamma}\Phi_\chi(\epsilon_0^\nu L) \,,
\end{align}
where $\beta$ and $\gamma$ are the critical exponents, and
$\Phi_{M/\chi}(z)$ are the universal scaling functions.
The scaling functions asymptotically approach
$\Phi_{M/\chi}(z)=1$ for $z\to\infty$ while
$\Phi_{M}(z)\sim z^{-\beta/\nu}$ and $\Phi_{\chi}(z)\sim z^{\gamma/\nu}$
for $z\to 0$. Here, for simplicity we have ignored the irrelevant difference in $\Delta_i = \Delta_f = \Delta$ above and below the critical point. Then the overlap between
the intervals $I_{i/f}^\delta$ is characterized by a single parameter,
$R\equiv M_i/\Delta$, the \emph{relative separation} between the two peaks of
$P_{i/f}(S)$.
Putting Eqs.~(\ref{Paper::eq:3}) and (\ref{Paper::eq:5}) together with the Rushbrooke scaling law,
$\alpha + 2\beta + \gamma = 2$  \cite{Goldenfeld92a}, one has the relative separation
\begin{equation}
R \sim L^{d/2}\epsilon_0^{(2-\alpha)/2}
\frac{\Phi_M(\epsilon_0^\nu L)}{\sqrt{\Phi_\chi(\epsilon_0^\nu L)}}.
\end{equation}
Using the Josephson hyperscaling law, $d\nu=2-\alpha$ \cite{Goldenfeld92a},
it is further reduced to
\begin{equation}
\label{Paper::eq:7}
R(\epsilon_0,L)
\sim (\epsilon_0^\nu L)^{d/2}
\frac{\Phi_M(\epsilon_0^\nu L)}{\sqrt{\Phi_\chi(\epsilon_0^\nu L)}} \,.
\end{equation}
It is remarkable that the relative separation
\begin{math}
R(\epsilon_0,L) = R(\epsilon_0^\nu L)
\end{math}
does not depend on $\epsilon_0$ and $L$ separately but is a universal function of only the combination $\epsilon_0^\nu L=L/\xi_0$, where $\xi_0\sim\epsilon_0^{-\nu}$ is the correlation length at $\epsilon=\epsilon_0$. This implies that $\avg{e^{-\sigma}}_\delta$ is also a universal function of $\epsilon_0^\nu L$ alone, which is indeed confirmed by the numerical results shown in Figs.~\ref{Paper::fig:3} (a) and \ref{Paper::fig:3} (b). Note that the hyperscaling law breaks down either in dimensions higher than the upper critical dimension $d_*=4$ or in the mean-field approximation. In such cases, where $\alpha=0$ and $\nu=1/2$, $\avg{e^{-\sigma}}_\delta$ is not necessarily a universal function of $\epsilon_0^\nu L$ in general.

For sufficiently large systems ($\epsilon_0^\nu L\gg 1$), one expects sharp distribution functions. 
Indeed, in this limit it follows that
\begin{equation}
\label{Paper::eq:11}
R(\epsilon_0^\nu L)
\sim (\epsilon_0^\nu L)^{d/2} 
\xrightarrow{L\to\infty} \infty
\end{equation}
and $P_{i/f}(S)$ are well separated.
On the other hand, when the system is small ($\epsilon_0^\nu L \ll 1$) and  finite-size effect sets in,
the larger fluctuations lead to broader distribution functions giving
\begin{equation}
\label{Paper::eq:12}
R(\epsilon_0^\nu L)
\sim (\epsilon_0^\nu L)^{(d\nu-2+\alpha)/2\nu}=1
\end{equation}
according to the hyperscaling law.
It means that $P_{i/f}(S)$ have significant overlap with each other for a finite-size system. 

With the universal scaling behaviors of relative separation $R$ at hand, let us now investigate $\avg{e^{-\sigma}}_\delta$ in the parameter space of $\epsilon_0^\nu L$ and $\delta$.
For a given tolerance $\delta$, the two asymptotic behaviors in Eqs.~(\ref{Paper::eq:11}) and (\ref{Paper::eq:12}) imply little and significant overlap between $P_{i/f}(S)$, respectively, and hence that $\avg{e^{-\sigma}}_\delta \sim 0$ in the limit of $\epsilon_0^\nu L\gg 1$ while $\avg{e^{-\sigma}}_\delta\sim1/2$ in the opposite limit of $\epsilon_0^\nu L\ll 1$; see Eq.~(\ref{Paper::eq:4}) and Fig.~\ref{Paper::fig:2}.
For $L$ fixed, on the other hand, large tolerance ($\delta\sim 1$) naturally leads to $\avg{e^{-\sigma}}_\delta\sim 0$ whereas we have seen above that $\avg{e^{-\sigma}}_\delta\sim1/2$ with $\delta\simeq 0$.
In short, as illustrated in Figs.~\ref{Paper::fig:3}(c) and (d), $\avg{e^{-\sigma}}_\delta\sim 0$ for $\epsilon_0^\nu L\gg 1$ and $\delta\sim 1$ while $\avg{e^{-\sigma}}_\delta\sim1/2$ for $\epsilon_0^\nu L\ll 1$ and $\delta\sim 0$.

Evidently, a crossover of $\langle e^{-\sigma}\rangle$ occurs as a result of combined effects of finite size and tolerance. One can locate the crossover boundary $\delta=\delta_*(\epsilon_0^\nu L)$ by identifying $\delta_*$ for given $\epsilon_0^\nu L$ as the maximum tolerance allowing for significant overlap between $\calI_{i/f}^\delta$. More specifically, $\delta_*$ is such that the lower end of the interval $\calI_i^\delta$ (i.e., $\min\calI_i^\delta$; recall that $M_i>0$) equals to the center (i.e., $M_f=0$) of $\calI_f^\delta$:
\begin{math}
\delta_*(\epsilon_0^\nu L)
\equiv 2\int_{-\infty}^0{dS}\,P_i(S).
\end{math}
The resulting crossover boundaries are illustrated by the thick red lines in Figs.~\ref{Paper::fig:3} (c) and \ref{Paper::fig:3}(d).

One can investigate $\avg{e^{-\sigma}}_\delta$ more closely for sufficiently large systems ($\epsilon_0^\nu L\gg 1$). In such a limit, the distributions $P_{i/f}(S)$ are sharp enough to be approximated by Gaussian forms. Then we observe (see Section I of \cite{Supplement}) that
\begin{math}
\delta_*
\sim (\epsilon_0^\nu L)^{-d/2}
e^{-(\epsilon_0^\nu L)^{d}/2}
\end{math}
and that
\begin{math}
\avg{e^{-\sigma}}_\delta
\sim (1-\delta)^{-1}\erfc^{-1}(\delta)
e^{-(\epsilon_0^\nu L)^d/2}.
\end{math}
In other words, in practice $\delta\gg\delta_*$ always and
$\avg{e^{-\sigma}}_\delta$ tends to vanish exponentially in the thermodynamic limit.

\paragraph{Conclusion and Discussions ---}
Taking the Ising model as an example, we studied the nonequilibrium process of driving across a second-order phase transition focusing on the deviation from the JE as a ``probe''. By introducing the tolerance parameter $\delta$, finite sampling of the initial ensemble was taken into account. We have found that  for a given $\delta$, the average of the exponential of the entropy production $\langle e^{-\sigma}\rangle$ is a universal function of $\epsilon_0^\nu L = L/\xi_0$.
As noted previously \cite{Zuckerman02a, Gore03a, Jarzynski06a, Palassini11a, Suarez12a,Lua05a, Sung05a, Gross05a, Jarzynski05a, Horowitz10a, Murashita14a}, the JE may break down for many practical and intrinsic reasons. Its breakdown for the dynamical processes of second-order phase transitions is peculiar as the deviation is determined by an universal combination of the system size $L$ and the initial reduced coupling $\epsilon_0$, which is inherited from the equilibrium scaling behavior of second-order phase transitions. It is stressed that such a universal scaling behavior is not limited to the sudden quenching but holds in general due to the critical slowing down (see Section II of \cite{Supplement}).
Our findings may provide a unique application of the Jarzynski equality to study the dynamical properties of phase transitions.

We thank Oscar Dahlsten and Carlo Danieli for helpful comments.
S.H. and M.-S.C. are supported by the National Research Foundation (NRF; Grant No.~2015-003689) and by the Ministry of Education of Korea through the BK21 Plus Initiative. D.-T.H., J.J., B.P.V., and G.W. are supported jointly by the Max Planck Society, the MSIP of Korea, Gyeongsangbuk-Do and Pohang City through the JRG at APCTP. G.W. also acknowledges support by the NRF (Grants No.~2012R1A1A2008028) and by the IBS (Grant No.~IBS-R024-D1).
J.J. and D.-T.H. also acknowledge support by the NRF (Grant No.~2013R1A1A1006655).

\bibliographystyle{apsrev}
\bibliography{Paper}

\begin{thebibliography}{34}
\bibitem[{end()}]{endnote:5}
\bibinfo{note}{In the presence of small tolerance, $\avg{e^{-\sigma}}_\delta$
  can slightly exceed $1/2$. This is an artifact due to the renormalization of
  the distribution function $P(S)$ truncated in accordance with the tolerance
  parameter.}
\end{thebibliography}

\end{document}